\documentstyle[a4,11pt]{article}

\date{}

\begin{document}

\title{{\bf Covariant approach to the conformal dynamical
equivalence in astrophysics}}

\author{Nicholas K. Spyrou$^1$ and Christos G. Tsagas$^{2,3}$\\
{\small $^1$~Astronomy Department, Aristoteleion University of
Thessaloniki}\\ {\small Thessaloniki 541.24, Macedonia, Hellas
(Greece)}\\ {\small $^2$~Department of Mathematics and Applied
Mathematics, University of Cape Town,}\\ {\small Rondebosch 7701,
South Africa}\\ {\small $^3$~DAMTP, Centre for Mathematical
Sciences, University of Cambridge, Cambridge CB3 0WA, UK}}

\maketitle

\begin{abstract}
We use covariant techniques to examine the implications of the
dynamical equivalence between geodesic motions and adiabatic
hydrodynamic flows. Assuming that the metrics of a geodesically
and a non-geodesically moving fluid are conformally related, we
calculate and compare their mass densities. The density difference
is then expressed in terms of the fundamental physical quantities
of the fluid, such as its energy and isotropic pressure. Both the
relativistic and the non-relativistic case are examined and their
differences identified. Our analysis suggests that observational
determinations of astrophysical masses based on purely Keplerian
motions could underestimate the available amount of
matter.
\end{abstract}

\section{Introduction}
Geodesic motions are central when it comes to mass measurements.
In particular, the observational determination of the masses of
various astrophysical objects is based on the assumption of
Keplerian, that is of purely geodesic, motions. For example, the
central mass concentration in various galaxies is determined by
Doppler-shift measurements of radiative sources which are assumed
to move along geodesic trajectories (e.g.~see~\cite{KR}).
Nevertheless, there are known cases where non-gravitational forces
are strong enough to affect the purely geodesic nature of these
trajectories and where a hydrodynamic description of the motion is
more appropriate~\cite{H}. In these cases one would like to know
whether the standard measurements overestimate or underestimate
the available amount of matter. This is the question the present
article will attempt to address.

We approach the problem by treating the geodesic and the
hydrodynamic descriptions of motion as dynamically
equivalent~\cite{KS}. Then, by introducing a {\em physical} fluid,
with an adiabatic hydrodynamic flow, and a {\em virtual} one, that
follows a geodesic trajectory, we can calculate and compare their
corresponding mass densities. Any difference found between the
virtual and the physical densities can be interpreted as a
correction to mass determinations based on the geodesic motion
approximation.\footnote{With the exception of Sec.~5, by mass we
will always imply the total energy of the matter. We will
therefore use the terms mass and energy interchangeably. In
Sec.~5, however, this is no longer the case and by mass we will
specifically indicate the rest mass of a Newtonian medium.}
In~\cite{KS} this was achieved by assuming that the metrics of the
two fluids are related through a simple conformal transformation
(see also~\cite{S} for further discussion). Moreover, when this
approach was applied to the central mass of certain Active
Galactic Nuclei (AGN), it showed that mass determinations based on
the geodesic motion approximation underestimated the available
amount of matter~\cite{KS}.

Here we provide a covariant generalisation of the analysis
introduced in~\cite{KS} that allows for a unified treatment of
both the Newtonian and the relativistic cases. Among the
advantages of the covariant formalism are its compactness and the
use of variables with a straightforward geometrical and physical
interpretation~\cite{E}. This facilitates a mathematically clear
and physically transparent treatment of the problem at hand. Our
analysis does not impose any a priori symmetries on the spacetime
metric and therefore applies to a range of physically realistic
situations. The only restrictions are on the nature of the medium,
which is assumed to be a perfect fluid, and the adiabaticity of
the flow. Following~\cite{KS}, we translate the dynamical
equivalence between the geodesic motion of the virtual fluid and
the hydrodynamic flow of its physical counterpart into a simple
conformal relation between their respective metrics. This allows
us to calculate the corresponding mass densities and express their
difference as a function of fundamental fluid variables, like its
energy density and pressure. As in~\cite{KS}, we interpret this
difference as a correction to mass determinations based on the
assumption of purely geodesic motions. Our analysis points towards
a mass deficit every time the geodesic motion approximation is
used for mass measurements.

\section{Energy difference between geodesic and hydrodynamic
flows}
Consider a spacetime (${\cal M},\,g_{\mu\nu}$) and the conformal
geometry of ${\cal M}$, namely the set of all metrics
$\tilde{g}_{\mu\nu}$ conformal to the physical metric $g_{\mu\nu}$
so that $\tilde{g}_{\mu\nu}=\Omega^2g_{\mu\nu}$, where $\Omega$ is
a nonzero suitably differentiable function.\footnote{Greek indices
run form 0 to 3, while Latin indices take the values 1,2,3. Also,
throughout this paper we employ a Lorenzian metric with signature
$(+\,-\,-\,-)$.} Under such transformations geodesic curves do not
generally remain geodesics unless they are null~\cite{HE}. On
these grounds, one could consider the adiabatic hydrodynamic flow
of a fluid in a given spacetime, and assume that it is represented
by the geodesic motion of a virtual fluid in another spacetime,
which is conformally related to the first. Then, adiabaticity and
the dynamical equivalence of the two motions specifies the
conformal factor to be~\cite{KS}
\begin{equation}
\Omega=\frac{{\cal E}+p}{\rho c^2}\,,  \label{Omega}
\end{equation}
where ${\cal E}=\rho c^2+\rho\Pi$ is the energy density of the
fluid, with $\rho$ being its rest-mass density and $\Pi$ the
internal energy per unit mass~\cite{KS,St}. According to
(\ref{Omega}), $\Omega\geq1$ for all conventional types of matter
with $p\geq0$. Note that the limit $\Omega=1$ corresponds to
``cold matter'' (i.e.~``dust'') with $p=0=\Pi$. Obviously, in that
case $\tilde{g}_{\mu\nu}=g_{\mu\nu}$ and the two spacetimes
coincide.

The aforementioned dynamical equivalence translates into a
relation between the variables of the two fluids, namely the
physical and the virtual one. In particular, if $\tilde{{\cal E}}$
is the energy density of the virtual fluid and ${\cal E}$ is that
of the physical fluid, then~\cite{KS}
\begin{equation}
\tilde{{\cal E}}=\Omega^{-2}({\cal E}+{\cal E}_1)\,.  \label{KS1}
\end{equation}
Therefore, ${\cal E}_1$ provides a measure of the energy density
difference that results from the dynamical equivalence of the two
motion representations. This difference, which will be interpreted
as a correction to any mass determination based on purely geodesic
trajectories, is given by
\begin{equation}
\Delta{\cal E}\equiv{\cal E}-\tilde{{\cal E}}=(1-\Omega^{-2}){\cal
E}- \Omega^{-2}{\cal E}_1\,,  \label{Delta}
\end{equation}
with the right-hand side expressed relative to the quantities of
the physical fluid. In particular, following~\cite{KS} we have
\begin{equation}
{\cal E}_1=-\frac{2}{\kappa}\frac{\Box\,\Omega}{\Omega}+
\frac{2}{\kappa}\frac{1}{\Omega}u^{\mu}u^{\nu}\nabla_{\nu}\nabla_{\mu}\Omega-
\frac{4}{\kappa}\frac{1}{\Omega^2}u^{\mu}u^{\nu}\nabla_{\mu}\Omega\nabla_{\nu}\Omega+
\frac{1}{\kappa}\frac{1}{\Omega^2}g^{\mu\nu}\nabla_{\mu}\Omega\nabla_{\nu}\Omega\,,
\label{KS2}
\end{equation}
where $u_{\mu}$ is the dimensionless timelike 4-velocity vector of
the matter normalised so that $u_{\mu}u^{\mu}=1$, $\nabla_{\mu}$
is the standard covariant derivative operator and $\kappa=8\pi
G/c^4$. As expected, both ${\cal E}_1$ and $\Delta{\cal E}$ vanish
at the $\Omega=1$ limit. Also, from Eq.~(\ref{Delta}) it becomes
clear that, since $\Omega>1$, $\Delta{\cal E}$ is positive if
${\cal E}_1$ is negative. Moreover, even when ${\cal E}_1>0$ the
energy difference will remain positive provided ${\cal
E}_1<(\Omega^2-1){\cal E}$, as Eq.~(\ref{Delta}) guarantees. In
any such case the energy density of the virtual fluid will be less
than that of the actual one. Whenever this happens mass
determination based on purely geodesic motions will underestimate
the matter content of the medium. In what follows we will
investigate this possibility in more detail while trying to
maintain the generality of our discussion.

\section{Covariant description of the energy difference}
Using $u_{\mu}$, the timelike 4-velocity vector of the matter, we
define $h_{\mu\nu}=g_{\mu\nu}-u_{\mu}u_{\nu}$ as our projection
tensor orthogonal to $u_{\mu}$. Then, the four terms in the
right-hand side of Eq.~(\ref{KS2}) are respectively recast as
\begin{eqnarray}
\Box\;\Omega\equiv g^{\mu\nu}\nabla_{\mu}\nabla_{\nu}\Omega&=&
{\rm D}^2\Omega+ \ddot{\Omega}- \dot{u}^{\mu}{\rm
D}_{\mu}\Omega\,, \label{rel1}\\
u^{\mu}u^{\nu}\nabla_{\nu}\nabla_{\mu}\Omega&=&
\ddot{\Omega}-\dot{u}^{\mu}{\rm
D}_{\mu}\Omega\,,  \label{rel2}\\
u^{\mu}u^{\nu}\nabla_{\mu}\Omega\nabla_{\nu}\Omega&=&
\dot{\Omega}^2\,,  \label{rel3}\\
g^{\mu\nu}\nabla_{\mu}\Omega\nabla_{\nu}\Omega&=&{\rm
D}_{\mu}\Omega\,{\rm D}^{\mu}\Omega+ \dot{\Omega}^2\,,
\label{rel4}
\end{eqnarray}
where $\dot{\Omega}=u^{\mu}\nabla_{\mu}\Omega$, ${\rm
D}_{\mu}\Omega=h_{\mu}{}^{\nu}\nabla_{\nu}\Omega$ and ${\rm
D}^2\Omega=h^{\mu\nu}\nabla_{\mu}\nabla_{\nu}\Omega$ by
definition, while $\dot{u}_{\mu}=u^{\nu}\nabla_{\nu}u_{\mu}$ is
the 4-acceleration. Note that an overdot indicates time
derivatives, that is covariant derivatives projected along the
timelike 4-velocity vector $u_{\mu}$. On the other hand, ${\rm
D}_{\mu}=h_{\mu}{}^{\nu}\nabla_{\nu}$ is the covariant derivative
operator orthogonal to $u_{\mu}$, which is used to denote the
local spatial gradients. Finally, ${\rm
D}^2=h^{\mu\nu}\nabla_{\mu}\nabla_{\nu}$ is the projected
Laplacian operator.

Substituting the results (\ref{rel1})-(\ref{rel4}) into
Eq.~(\ref{KS2}) we arrive at the following alternative, covariant
expression for the key quantity ${\cal E}_1$
\begin{equation}
{\cal E}_1=-\frac{1}{\kappa}\left(\frac{2}{\Omega}\;{\rm
D}^2\Omega+ \frac{3}{\Omega^2}\;\dot{\Omega}^2-
\frac{1}{\Omega^2}\;{\rm D}_{\mu}\Omega\,{\rm
D}^{\mu}\Omega\right)\,. \label{delta1}
\end{equation}
Of the three terms on the right-hand side, the second is negative
definite while the third is positive definite. The former of these
two terms suggests that mass determinations based on the geodesic
motion approximation will underestimate the total energy density
of the matter, whereas the latter indicates the opposite. The sign
of the first term, on the other hand, is not a priori fixed but
depends on the sign of the spatial Laplacian ${\rm D}^2\Omega$.
Clearly, the overall effect of these three terms depends on their
relative strength. Note that the third term in Eq.~(\ref{delta1})
is quadratic in ${\rm D}_{\mu}\Omega$, which suggests that it
should become strong only in highly inhomogeneous situations.

\section{Energy difference and the fluid parameters}
For a better insight into the physical implications of the
equivalence between geodesic and hydrodynamic motions, it helps to
recast expression (\ref{delta1}) with respect to the fundamental
physical quantities of the fluid, namely $\rho$, $p$ and ${\cal
E}$. As before, we will do so by restricting our analysis to
isentropic (i.e.~adiabatic) flows, which means that
\begin{equation}
\nabla_{\mu}\Pi=\frac{p}{\rho^2}\nabla_{\mu}\rho\,.
\label{0entropy}
\end{equation}
without any extra constraints on the metric~\cite{KS,S}. Then,
taking the covariant derivative of Eq.~(\ref{Omega}) and using
result (\ref{0entropy}) we obtain
\begin{equation}
\nabla_{\mu}\Omega=\frac{1}{\rho c^2}\nabla_{\mu}p\,.
\label{nablaOmega}
\end{equation}
Contracted along the timelike direction $u_{\mu}$ and projected
orthogonal to $u_{\mu}$ the above gives
\begin{equation}
\dot{\Omega}=\frac{\dot{p}}{\rho c^2}\,  \hspace{10mm} {\rm and}
\hspace{10mm} {\rm D}_{\mu}\Omega=\frac{1}{\rho c^2}\;{\rm
D}_{\mu}p\,, \label{grad-dotOmega}
\end{equation}
respectively. Moreover, the latter of these two results leads to
\begin{equation}
{\rm D}^2\Omega=\frac{1}{\rho c^2}\;{\rm D}^2p- \frac{1}{\rho^2
c^2}\;{\rm D}_{\mu}\rho\,{\rm D}^{\mu}p\,. \label{laplOmega}
\end{equation}
On using (\ref{grad-dotOmega}) and (\ref{laplOmega}) we may recast
expression (\ref{delta1}) as
\begin{equation}
{\cal E}_1=-\frac{1}{\kappa\,\Omega\rho c^2}\left(2{\rm D}^2p+
\frac{3}{\Omega\rho c^2}\;\dot{p}^2- \frac{2}{\rho}\;{\rm
D}_{\mu}\rho{\rm D}^{\mu}p- \frac{1}{\Omega\rho c^2}\;{\rm
D}_{\mu}p{\rm D}^{\mu}p\right)\,, \label{delta2}
\end{equation}
where $\Omega\rho c^2={\cal E}+p$ (see Eq.~(\ref{Omega})). The
above expresses the difference, between the observationally
determined (i.e.~the virtual) and the actual energy density, as a
function of key physical quantities like the fluid rest-mass
density and isotropic pressure. Moreover, it shows that both
${\cal E}_1$ and $\Delta{\cal E}$ vanish when the pressure is
zero. As expected, this means that the use of geodesic motions
will lead to a noticeable underestimation (or overestimation) of
the available mass mainly in domains where the fluid pressure and
its gradients are appreciable. Such domains could be the highly
dense central regions of certain active galaxies.

Expression (\ref{delta2}) also provides a physical interpretation
to some of the mathematical comments made immediately after
Eq.~(\ref{delta1}). For example, the sign of the Laplacian term in
(\ref{delta1}) depends on whether ${\rm D}^2p$ is positive or
negative. This, in turn, depends on whether we are looking at an
overdense or an underdense region respectively. In other words,
the Laplacian term, alone, will underestimate the fluid energy
density whenever the geodesic motion approximation is applied to
an overdense region. Note that for our purposes we will always
consider overdense regions, which means that the Laplacian term in
the right-hand side of (\ref{delta2}) will always be treated as
positive. Also, according to (\ref{delta2}), time variations in
the fluid pressure (and therefore in its energy density) will also
underestimate the available energy. Spatial gradients in the fluid
pressure, on the other hand, could lead to an energy
overestimation. This latter effect, which depends on the specific
nature of the fluid, will be addressed in more detail in the
following sections.

\section{The case of a Newtonian medium}
Equation (\ref{delta2}) applies to a general relativistic fluid in
strong gravity environments. Most astrophysical situations,
however, are adequately described by the Newtonian theory. In this
respect, it is worth recovering the classical limit of
Eq.~(\ref{delta2}). We will do that next for the case of a
polytropic gas with $p=k\rho^{\gamma}$, where the parameters $k$
and $\gamma$ will be treated as constants. For such a medium we
have
\begin{equation}
{\cal E}_1=-\frac{v_{\rm s}^2}{\kappa({\cal E}+p)}\left[2{\rm
D}^2\rho +\frac{1}{\rho}\left(2(\gamma-2)-\frac{v_{\rm
s}^2\rho}{{\cal E}+p}\right){\rm D}_{\mu}\rho{\rm D}^{\mu}\rho
+\frac{3v_{\rm s}^2}{{\cal E}+p}\,\dot{\rho}^2\right]\,,
\label{poldelta}
\end{equation}
where $v_{\rm s}^2={\rm d}p/{\rm d}\rho=\gamma p/\rho$ is the
square of the (adiabatic) polytropic sound speed. For a
non-relativistic gas, we have $\Pi/c^2\ll1$ and therefore ${\cal
E}\simeq\rho c^2$. Also, $p\ll\rho c^2$ implying that ${\cal
E}+p\simeq\rho c^2$ and that $v_{\rm s}^2/c^2\ll1$. At this limit
the above equation reduces to
\begin{equation}
\rho_1\simeq-\frac{2v_{\rm s}^2}{\kappa\rho c^4}\left[{\rm
D}^2\rho +\frac{\gamma-2}{\rho}\;{\rm D}_{\mu}\rho{\rm
D}^{\mu}\rho\right]\,, \label{n-rpoldelta}
\end{equation}
having dropped all terms quadratic in $p/\rho c^2$ and $v_{\rm
s}^2/c^2$. Note that for a Newtonian fluid expression
(\ref{Delta}) reduces to $\Delta\rho\simeq-\rho_1$ to lowest order
in $p/\rho c^2$ and $v_{\rm s}^2/c^2$. In addition, at the weak
gravity limit we can replace the projected covariant derivative
operators with ordinary partial derivatives (i.e.~${\rm
D}_{\mu}\rightarrow\partial_a$). Then,
\begin{equation}
\rho_1\simeq-\frac{k\gamma\rho^{\gamma-2}}{4\pi
G}\left[\partial^2\rho
+\frac{\gamma-2}{\rho}\;\partial_a\rho\partial^a\rho\right]\,,
\label{Npoldelta}
\end{equation}
where $v_{\rm s}^2=k\gamma\rho^{\gamma-1}$ is the sound speed of
our polytropic gas and $\partial^2\equiv\partial_a\partial^a$ is
the standard 3-dimensional Laplacian. The above equation, which
has been derived by reducing the fully relativistic covariant
formula (\ref{delta2}) to the Newtonian limit, agrees completely
with the result obtained in~\cite{KS} via a metric-based approach.
Following result (\ref{Npoldelta}) we conclude that $\rho_1<0$,
and therefore $\Delta\rho>0$, whenever $\gamma\geq2$. Note that,
due to the Laplacian term in the right-hand side of
Eq.~(\ref{Npoldelta}), which is assumed positive, the
aforementioned condition on the polytropic index for $\Delta\rho$
to be positive is sufficient but not necessary. In practise, this
means that $\Delta\rho>0$ for almost all matter distributions
where the pressure increases with the density. Moreover, the
$\gamma$-dependence effectively disappears when the inhomogeneity
is weak and the second term in the brackets becomes negligible.
Based on these results one may argue that determining the mass
content in a `Newtonian region' on the assumption of geodesic
motions will generally underestimate the available amount of
matter.

\section{The case of a relativistic medium}
Let us now turn our attention to a relativistic fluid and use
expression (\ref{delta2}) to identify the differences from the
above treated Newtonian case. To begin with, for a perfect fluid
the standard energy density conservation means that
\begin{equation}
\dot{\cal E}=-(1+w)\Theta{\cal E}\,,  \label{edc}
\end{equation}
where $w=p/{\cal E}$ and $\Theta=\nabla_{\mu}u^{\mu}={\rm
D}_{\mu}u^{\mu}$. Note that the scalar $\Theta$ describes the
volume expansion (or contraction) between the worldlines of two
neighbouring fluid particles (e.g.~see~\cite{E}). Moreover, the
barotropic equation of state of the fluid, namely the fact that
the pressure is a function of the energy density alone
(i.e.~$p=p({\cal E})$), implies that
\begin{equation}
{\rm D}_{\mu}p=c_{\rm s}^2{\rm D}_{\mu}{\cal E} \hspace{10mm} {\rm
and} \hspace{10mm} \dot{p}=c_{\rm s}^2\dot{\cal E}\,,  \label{bar}
\end{equation}
where $c_{\rm s}^2={\rm d}p/{\rm d}{\cal E}$ is the dimensionless
adiabatic sound speed. Then, assuming that the equation of state
of the fluid does not change (i.e.~setting $w$, $c_{\rm s}^2=$
constant) in the region under consideration, we obtain
\begin{equation}
{\rm D}^2p=c_{\rm s}^2{\rm D}^2{\cal E}\,.  \label{bar1}
\end{equation}
Finally, in addition to constraint (\ref{0entropy}), adiabaticity
also guarantees that
\begin{equation}
\frac{1}{{\cal E}(1+w)}\nabla_{\mu}{\cal E}=
\frac{1}{\rho}\nabla_{\mu}{\rho}\,. \label{0entropy1}
\end{equation}
Note that, when the parameters $w$, $c_{\rm s}^2$ are not
constant, one needs to specify the equation of state of the fluid.

On using results (\ref{edc})-(\ref{0entropy1}), expression
(\ref{delta2}) reads
\begin{equation}
{\cal E}_1=-\frac{c_{\rm s}^2}{\kappa{\cal
E}^2(1+w)^2}\left[2{\cal E}(1+w){\rm D}^2{\cal E}+ 3c_{\rm
s}^2(1+w)^2\Theta^2{\cal E}^2- (2+c_{\rm s}^2){\rm D}_{\mu}{\cal
E}{\rm D}^{\mu}{\cal E}\right]\,.  \label{DeltacE1}
\end{equation}
Accordingly ${\cal E}_1=0$ when $c_{\rm s}^2=0$. The latter
implies that the fluid pressure is a covariantly constant quantity
(i.e.~isobaric flow with $\nabla_{\mu}p=0$ - see
Eqs.~(\ref{bar})). Then, for a barotropic fluid, isobaric flow
means that ${\cal E}$ is also covariantly constant. We will return
to the implications of Eq.~(\ref{DeltacE1}) for a relativistic
medium in the next section. For the moment, we simply point out
that the sign of the right-hand side of (\ref{DeltacE1}) depends
on the relative contribution of the three terms in the brackets.

Note that when ${\rm D}_{\mu}{\cal E}=0$ but $\dot{{\cal
E}}\neq0$, namely for a fluid with a spatially homogeneous energy
density distribution, expression (\ref{DeltacE1}) reduces to
${\cal E}_1=-3c_{\rm s}^4\Theta^2/\kappa<0$. In this case we have
a positive energy difference (i.e.~$\Delta{\cal E}>0$ - see
Eq.~(\ref{KS1})), which implies that mass calculations based on
geodesic motions will lead to an energy deficit. Interestingly,
this last result is independent of whether there is expansion or
contraction, but vanishes when we consider a stationary region.

\section{The energy deficit}
The Newtonian case was addressed in~\cite{KS} by considering a
spherically symmetric region and assuming an outwardly decreasing
mass density distribution of the Plummer-type
\begin{equation}
\rho=\rho_0\left[1+\left(r/r_0\right)^2\right]^{-n/2}\,,
\label{Plummer}
\end{equation}
where $r$ is the radius of the region in question, $\rho_0$ and
$r_0$ are constants and $n$ is a positive integer with typical values
either 3 or 5 (e.g.~see~\cite{BT}). Note that the
Plummer parameters are related by $\rho_0=2^{n/2}\rho(r_0)$, an
expression that will prove useful later in this section. For
$\gamma\leq1+1/n$ it can be shown that $\rho_1$, as given by
(\ref{Npoldelta}), is always negative which implies
that the geodesic motion approximation will underestimate the
available matter density in the region~\cite{KS}.

In the Newtonian limit one could reasonably well ignore terms
quadratic in $p/\rho c^2$ and $v_{\rm s}^2/c^2$, given that the
pressure is a negligible fraction of the matter density. This
means that we can disregard the contribution of the last two terms
in Eq.~(\ref{poldelta}). When dealing with a relativistic medium,
however, this is no longer an option, since the pressure forms a
considerable fraction of the matter-energy density. In this case
one should include the last two terms in Eq.~(\ref{DeltacE1}).
Their relative contribution depends on whether the region is
stationary or not, and also on the degree of the inhomogeneity. In
principle, one could apply expression (\ref{DeltacE1}) to any
situation and to any given density distribution, and then
calculate the energy deficit or surplus due to the geodesic motion
approximation. However, the complexity of the relativistic
equations means that analytic quantitative results are very
difficult to extract, even for the relatively simple Plummer
distribution. Nevertheless, one can still obtain some general
qualitative results and this is what we will try to do next.

As mentioned before, the last term in the right-hand side of
Eq.~(\ref{DeltacE1}) will become important only in highly
inhomogeneous situations. Indeed, assuming Euclidean geometry for
simplicity, a dimensional analysis shows that
\begin{equation}
{\rm D}^2{\cal E}\sim\frac{\delta{\cal E}}{\delta\lambda^2}\,
\hspace{10mm} {\rm and} \hspace{10mm}  {\rm D}_{\mu}{\cal E}{\rm
D}^{\mu}{\cal E}\sim \left(\frac{\delta{\cal
E}}{\delta\lambda}\right)^2\,, \label{approx}
\end{equation}
where ${\cal E}$ is the average energy density of the medium and
$\delta\lambda$ is the characteristic length scale associated with
the spatial variation of ${\cal E}$. So, as long as the density
contrast $\delta=\delta{\cal E}/{\cal E}$ is small compared to
unity, we have
\begin{equation}
\frac{1}{{\cal E}}{\rm D}^2{\cal E}\sim \frac{\delta{\cal E}/{\cal
E}}{\delta\lambda^2}\gg \left(\frac{\delta{\cal E}/{\cal
E}}{\delta\lambda}\right)^2\sim \frac{1}{{\cal E}^2}{\rm
D}_{\mu}{\cal E}{\rm D}^{\mu}{\cal E}\,,
\end{equation}
and the Laplacian term in the right-hand side of (\ref{DeltacE1})
dominates over the quadratic term ${\rm D}_{\mu}{\cal E}{\rm
D}^{\mu}{\cal E}$. Then,
\begin{equation}
{\cal E}_1\simeq-\frac{c_{\rm s}^2}{\kappa{\cal E}(1+w)}
\left[2{\rm D}^2{\cal E}+ 3c_{\rm s}^2(1+w)\Theta^2{\cal
E}\right]<0\,. \label{domDeltacE1}
\end{equation}
where the last term contributes only in non-stationary regions.
Therefore, in a weakly inhomogeneous environment, the use of
geodesic motions for the observational determination of the matter
content will underestimate the available energy density in the
region.

In highly inhomogeneous situations, however, $\delta>1$ and the
same dimensional argument shows that this time it is the quadratic
term that dominates over the Laplacian in the right-hand side of
(\ref{DeltacE1}). In this case,
\begin{equation}
{\cal E}_1\simeq-\frac{c_{\rm s}^2}{\kappa{\cal
E}^2(1+w)^2}\left[3c_{\rm s}^2(1+w)^2\Theta^2{\cal E}^2-(2+c_{\rm
s}^2){\rm D}_a{\cal E}{\rm D}^a{\cal E}\right]\,.
\label{domDeltacE2}
\end{equation}
Since both $w$ and $c_{\rm s}^2$ are of order unity, the sign of
the above depends primarily on two parameters. These are the
contraction scalar $\Theta$, which provides a measure of the
average collapse timescale, and the ratio ${\rm D}_a{\cal E}/{\cal
E}\sim\delta/\delta\lambda$, which determines the degree
($\delta$) and the scale ($\delta\lambda$) of the inhomogeneity.
Therefore, provided the collapse timescale is short enough, that
is as long as $1<\delta<\Theta\delta\lambda$, we will have ${\cal
E}_1<0$ and a deficit in the estimated energy density
(i.e.~$\Delta{\cal E}>0$). It should be emphasised that
$\delta>\Theta\delta\lambda$ does not automatically imply a
negative energy difference and an overestimated energy density.
Indeed, suppose that the region in question is stationary. In that
case the right-hand side of (\ref{domDeltacE2}) is dominated by
the inhomogeneous term and
\begin{equation}
{\cal E}_1\simeq\frac{c_{\rm s}^2(2+c_{\rm s}^2)}{\kappa{\cal
E}^2(1+w)^2}\,{\rm D}_{\mu}{\cal E}{\rm D}^{\mu}{\cal E}>0\,.
\label{domDeltacE3}
\end{equation}
Following (\ref{Delta}), the above leads to $\Delta{\cal E}<0$
only if
\begin{equation}
\frac{1}{{\cal E}^2}\;{\rm D}_{\mu}{\cal E}{\rm D}^{\mu}{\cal E}>
\frac{\kappa(\Omega^2-1)(1+w)^2}{c_{\rm s}^2(2+c_{\rm s}^2)}\;
{\cal E}\,,  \label{surplus}
\end{equation}
which gives a measure of the inhomogeneity required for a negative
energy density difference. Given that $c_{\rm s}^2(2+c_{\rm
s}^2)/(\Omega^2-1)(1+w)^2$ is of order unity, the above translates
into the following order of magnitude condition on ${\cal E}$,
$\delta$ and $\delta\lambda$
\begin{equation}
\delta>\sqrt{\kappa{\cal E}}\delta\lambda\,.  \label{surplus1}
\end{equation}
for $\Delta{\cal E}<0$. In other words, mass determinations based
on the assumption of geodesic motions in a region filled with a
highly inhomogeneous relativistic medium, will underestimate the
available amount of energy unless condition (\ref{surplus1}) is
satisfied. It is therefore interesting to examine whether the
above condition is typically satisfied or not. Assuming a
Plummer-type density distribution for the relativistic fluid, then
in absolute values the associated density contrast is
$\delta=nx^2(\delta r/r)/(1+x^2)$, with $x=r/r_0$ by definition.
In this environment, condition (\ref{surplus1}) becomes
\begin{equation}
x(1+x^2)^{\frac{n}{4}-1}>\frac{r_0}{nc}\sqrt{8\pi G\rho_0}\,,
\label{rPlummer1}
\end{equation}
since ${\cal E}=\rho c^2$ and $\delta\lambda=\delta r$ given the
symmetries of the Plummer distribution. Then, for $x\gg1$ the
above translates into the following condition involving the Plummer
parameters
\begin{equation}
\frac{r_0}{nc}\sqrt{8\pi G\rho_0}\gg1\,,  \label{rPlummer2}
\end{equation}
where $n>2$ (with typical values $n=3,\,5$)~\cite{M}. Consider now
a supergiant elliptical galaxy, with mass $M$ of the order of
$10^{13}-10^{14}~M_{\odot}$, and assume that $r_0=3R_{\rm S}$,
where $R_{\rm S}=2GM/c^2$ is the Schwarzchild radius of the
central black hole. In that case
$\rho(r_0)\sim10^{-17}~gr/cm^{-3}$~\cite{SV}, which provides a
value for $\rho_0$ (recall that $\rho_0=2^{n/2}\rho(r_0)$ for a
Plummer-type density profile). Substituting the above into
condition (\ref{rPlummer2}) we obtain
\begin{equation}
M\gg n2^{-n/2}\times 8.261\times10^{15}~M_{\odot}\,,  \label{Mcon}
\end{equation}
imposing a lower bound of approximately $10^{16}~M_{\odot}$ on the
mass of the central compact object. Such a value is clearly way
above any accepted mass estimate for the core region of a galaxy,
which means that our original assumption that condition
(\ref{surplus1}) holds is false. In other words, as in the
Newtonian case, one can argue that mass determinations based on
purely geodesic motions will generally underestimate the energy
content in `relativistic regions' as well.

\section{Discussion}
Estimating the masses of the various astrophysical bodies is
central to almost all astronomical problems. Such mass
determinations use observational techniques which are based on the
assumption of geodesic motions, that is motions under the effect
of gravity alone. In the presence of non-gravitational forces,
however, due to, say, pressure gradients, viscosity or an
inhomogeneous magnetic field, a purely geodesic motion is no
longer sustainable. Then, a hydrodynamical description of the
motion is more appropriate, as it appears to be the case near the
core regions of typical AGNs. Therefore, the emerging question is
how accurate the standard mass measurements are and in particular
whether they underestimate or overestimate the available amount of
matter. Answering the question to the full is not an easy task and
is further complicated by a number of unknown parameters. The
latter are related to the nature of the astrophysical medium in
the region under consideration, its distribution, etc.
Nevertheless, one could still try to address the problem in a
series of qualitative steps. The first is to investigate whether,
and under what conditions, the standard methods underestimate or
overestimate the available amount of matter. Here, we have
attempted to address this issue by generalising the work
of~\cite{KS}. At the core of our approach is the conformal
dynamical equivalence between the motions of a virtual fluid,
which follows a geodesic trajectory, and of its physical
counterpart, which has an adiabatic hydrodynamical flow. The
assumption of adiabaticity is typical in analytical studies, which
by nature cannot address highly complicated physical systems. In
our case adiabaticity means that we consider systems which are
isolated enough to maintain their thermodynamical content
unchanged or slowly varying in time.

The dynamical equivalence between geodesic and hydrodynamic
motions translates into a simple conformal relation between the
metrics of the respective fluids. This in turn provides a
mathematical framework where one can calculate and compare the
matter content of the two fluids. Any difference found could then
be interpreted as a correction to mass determinations based on the
geodesic motion approximation. Based on that one can then proceed
to examine whether the assumption of geodesic motions
underestimates, or overestimates, the amount of matter available
in certain astrophysical formations. Here, we have considered the
conformal dynamical equivalence between geodesic motions and
hydrodynamical flows due to non-zero pressure gradients. However,
the same principle could be used to analyse more general
situations. One should be able to consider, for example, the
conformal dynamical equivalence between geodesic motions and
magneto-hydrodynamical flows.

Our approach offers a unified and fully covariant treatment of
both the Newtonian and the relativistic cases. The first task was
to provide the general mathematical framework where one can select
and study specific individual cases. For this reason we did not
impose any a priori constraints on the spacetime geometry, which
made our approach applicable to a range of physically realistic
situations. The second objective was to investigate whether or not
mass measurements using purely geodesic motions underestimate the
available amount of matter. According to our analysis, mass
determinations based on geodesic motions in a Newtonian
environment will underestimate the available matter of a
polytropic fluid, provided the associated polytropic index
satisfies the constraint $\gamma\geq2$. Crucially, this
$\gamma$-dependence can only become important in highly
inhomogeneous situations. The relativistic analysis also led to
similar conclusions. We found, in particular, that measurements
based on purely Keplerian motions will underestimate the available
amount of energy unless the region in question is both highly
inhomogeneous and stationary. Moreover, the latter possibility
appears unattainable in practise, at least for Plummer-type matter
distributions. Note that the Plummer-type profile, and others like
e.g.~the Navarro-Frenk-White and the Hernquist profiles, are
particular cases of a more general "universal" density
distribution, which are supported by observations, and, so, they
have been widely applied in the literature~\cite{He}-\cite{SW}.
Thus, based on the conformal dynamical equivalence scenario, we
feel confident enough to argue that in the majority of
astrophysical situations the use of geodesic motions for the
observational determination of masses will underestimate the
available amount of matter. Given that, one may also want to
consider the potential implications of the dynamical equivalence
principle for the large-scale properties of the universe. In the
present article, however, we have focused on astrophysical
situations and therefore we refer the reader to~\cite{S2,S3} for a
discussion on the possible global applications of this approach.
In astrophysics the questions we would like to address further are
whether the mass underestimation identified here for Plummer-type
density profiles is typical, and whether there are cases where a
gross underestimation of the available matter can occur. If it so
proves, it could force a radical revision of the current views on
issues as important as the amount of baryonic matter in our
universe and its present distribution.

\section*{Acknowledgements}
The authors would like to thank George Ellis for helpful comments.
CGT was supported by a Sida/NRF grant.

\end{document}